\documentclass[usenatbib,usegraphicx]{mn2e}

\usepackage{graphicx}

\usepackage{times}
\usepackage{aas_macros}
\usepackage{pdfpages}
\usepackage{grffile}
\usepackage{amsmath}
\usepackage{amsfonts}
\usepackage{booktabs}
\usepackage{bm}
\usepackage{arydshln}
\usepackage[pdftex,colorlinks=true,breaklinks=true,citecolor=blue,draft=false]{hyperref}

\date{Accepted ---. Received ---; in original form ---}
\pagerange{\pageref{sec:intro}--\pageref{lastpage}}
\pubyear{2019}
\begin{document}

\title[Load balancing for interferometric image reconstruction]{Load balancing for distributed interferometric image reconstruction}
\author[Pratley \& McEwen]
{Luke Pratley$^1$\thanks{E-mail: Luke.Pratley@gmail.com}, Jason D.~McEwen$^1$ \newauthor\\
${}^1$Mullard Space Science Laboratory (MSSL), University College London (UCL), Holmbury St Mary, Surrey RH5 6NT, UK\\}
\maketitle

\begin{abstract}
  We present a new algorithm to perform wide-field radio interferometric image reconstruction, with exact non-coplanar correction, that scales to big-data.  This algorithm allows us to image 2 billion visibilities on 50 nodes of a computing cluster for a 25 by 25 degree field of view, in a little over an hour. We build on the recently developed distributed $w$-stacking $w$-projection hybrid algorithm, extending it to include a new distributed degridding algorithm that balances the computational load of the $w$-projection gridding kernels.  The implementation of our algorithm is made publicly available in the PURIFY software package.  Wide-field image reconstruction for data sets of this size cannot be performed effectively using the allocated computational resources without computational load balancing, demonstrating that our algorithms are critical for next-generation wide-field radio interferometers. 
\end{abstract}


\section{INTRODUCTION}
\label{sec:intro}
Next-generation low frequency wide-field of view telescopes, such as the Murchison Widefield Array (MWA; \citealt{tin13}), have non-coplanar baselines and other instrumental effects that need to be modeled during image reconstruction. Furthermore, the large volumes of visibilities and large image sizes increase the computational burden of imaging observations from next-generation telescopes. However, there are major computational and instrumental challenges that need to be overcome for these telescopes to reach the high resolution and sensitivity required by science goals of next-generation telescopes, such as detection of the epoch of reionization (EoR) \citep{koo15} and to probe Galactic and extra-galactic magnetic fields. 

Recent novel developments in fast construction of $w$-projection kernels and the distributed $w$-stacking $w$-projection hybrid algorithm  \citep{cor08b,pra19b} has allowed fast and accurate modeling of non-coplanar effects over extremely wide-fields of view from the MWA for over 100 million measurements \citep{pra19d}. The algorithm allows parallel construction of $w$-projection kernels while also distributing their storage for application, proving to be an effective method of tackling the most computational and memory intensive components of radio interferometric imaging \citep{wor16,bra16,hol17,pra19b}. However, while this distribution reduces the size and computational cost of the $w$-projection kernel, it does not ensure that computational resources are being used most effectively across the compute cluster. This makes it vulnerable to bottlenecks in computation without the modifications presented in this work.

This work presents a new distributed gridding algorithm that evenly balances the computational load across a computing cluster, extending the distributed gridding methods developed in \citet{pra19c}. Such an approach allows full memory and computational use across the nodes of the computing cluster when performing fast Fourier transforms (FFTs) of $w$-stacks and when degridding with $w$-projection kernels, which has not been possible previously, removing resource bottlenecks when imaging wide-fields of view for large data sets.  Such distributed degridding and gridding algorithms will be vital for next-generation radio interferometers with large data sets, such as the Square Kilometer Array (SKA). In particular, such an algorithm is needed for effectively correcting instrumental effects via the image and Fourier domain, while using the full performance of a computing cluster.

The remaining sections of this article are as follows. Section \ref{sec:radio_measurement_equation} introduces the wide-field interferometric measurement equation and the distributed $w$-stacking $w$-projection hybrid algorithm. Section \ref{sec:bottleneck} discusses the computational and memory bottlenecks of the distribution method. Section \ref{sec:mo_dist_grid} presents the new algorithm that evenly distributes the computational load across compute nodes.  Section \ref{sec:impl}  demonstrates the application of this algorithm that has been implemented in the interferometric imaging software package PURIFY\footnote{\url{https://github.com/astro-informatics/purify}}.

\section{Wide-field Imaging Measurement Equation}					
\label{sec:radio_measurement_equation}
The interferometric measurement equation is a result of the van Cittert-Zernike theorem \citep{zer38} and it can be extended to include many aspects of the measurement process \citep{smi11}. One simplified variation is the non-coplanar wide-field interferometric measurement equation, it reads
\begin{equation}
\begin{split}
	y(u, v, w^\prime) = \int x(l, m) a(l, m)\frac{{\rm e}^{-2\pi i w^\prime(\sqrt{1 -l^2 - m^2} - 1)}}{\sqrt{1 -l^2 - m^2}}\\
	\times {\rm e}^{-2\pi i (lu + mv)}\,  {\rm d}l{\rm d}m\, ,
	\end{split}
	\label{eq:meas_eq}
\end{equation}
where $(u, v, w^\prime)$ are the baseline coordinates and $(l, m, n)$ are directional cosines restricted to the unit sphere. In this work, we define $w^\prime = w + \bar{w}$, where $\bar{w}$ is the average value of $w$-terms, and $w$ is the effective $w$-component (with zero mean),
$x$ is the sky brightness and $a$ includes direction dependent effects such as the primary beam. The measurement equation allows one to calculate model measurements $y$ when provided with a sky model $x$. 

A number of methods make use of the measurement equation to recover an image $x$ given visibilities $y$. Two examples in radio astronomy are CLEAN \citep{hog74,pra16} and Sparse Regularization algorithms \citep{mce10,ono16,LP18,da18,pra19b,pra19c}. 
																				
To make use of the FFT the measurement equation is traditionally calculated and approximated using degridding \citep{tho08}. The measurement equation can be represented by the following linear operations 
\begin{equation}
	\bm{y} = \bm{\mathsf{W}}\bm{\mathsf{[GC]}}\bm{\mathsf{F}}\bm{\mathsf{Z}}\bm{\mathsf{S}} \bm{x}\, .
	\label{eq:matrix_equation}
\end{equation}
$\bm{\mathsf{S}}$ represents a gridding correction and correction of baseline independent effects such as $\bar{w}$, $\bm{\mathsf{Z}}$ represents zero padding to increase resolution of the Fourier grid, $\bm{\mathsf{F}}$ is an FFT, $\bm{\mathsf{G}}$ represents a sparse circular convolution matrix that interpolates measurements off the grid, while $\bm{\mathsf{[GC]}}$ corrects baseline dependent effects and interpolates measurements off the grid, and $\bm{\mathsf{W}}$ are weights applied to the measurements. This linear operator is typically called a measurement operator $\bm{\mathsf{\Phi}} = \bm{\mathsf{W}}\bm{\mathsf{G}}\bm{\mathsf{C}}\bm{\mathsf{F}}\bm{\mathsf{Z}}\bm{\mathsf{S}}$ with $\bm{\mathsf{\Phi}} \in \mathbb{C}^{M \times N}$. Furthermore, $\bm{x}_i = x(\bm{l}_i)$ and $\bm{y}_q = y(\bm{u}_q)$ are discrete vectors in $\mathbb{R}^{N \times 1}$ and $\mathbb{C}^{M \times 1}$ in this setting. The dirty map can be calculated from the adjoint operation $\bm{\mathsf{\Phi}}^\dagger\bm{y}$, and the residual map by $\bm{\mathsf{\Phi}}^\dagger(\bm{\mathsf{\Phi}}\bm{x} - \bm{y})$. 

\subsection{Distributed wide-field measurement operator}
In the distributed $w$-stacking $w$-projection algorithm \citep{pra19b}, the measurement operator corrects for the average $w$-value in each $w$-stack, then applies an extra correction to each visibility with the $w$-projection. Each $w$-stack $\bm{y}_k$ has the measurement operator of
\begin{equation}
	\bm{\mathsf{\Phi}}_k = \bm{\mathsf{W}}_k\bm{\mathsf{[GC]}}_k\bm{\mathsf{F}}\bm{\mathsf{Z}}\bm{\mathsf{\tilde{S}}}_k\,.
\end{equation}
The gridding correction, $\bm{\mathsf{\tilde{S}}}_k$, has been modified to correct for the $w$-stack dependent effects, such as the average $w$-value of the stack $\bar{w}_k$
\begin{equation}
	{\bm{\mathsf{\tilde{S}}}_k}_{ii} = \frac{a_k(l_i, m_i){\rm e}^{-2\pi i \bar{w}_k(\sqrt{1 -l^2_i -m^2_i} - 1)}}{g(l^2_i + m^2_i)\sqrt{1 -l^2_i - m^2_i}}\, .
\end{equation}
We leave the option of choosing different primary beam effects in a stack $a_k(l_i, m_i)$. The chirp shifts the relative $w$-value in the stack indexed by $k$. The stacks can be clustered carefully to reduce the effective $w$-value in the stack, especially when the stack is close to the mean $\bar{w}_k$, i.e.\ to the value of $w_i - \bar{w}_k$. This reduces the size of the support needed in the $w$-projection gridding kernel for each stack,
\begin{equation}
	\begin{split}
		{\bm{\mathsf{[GC]}}_k}_{ip} =
		[GC]\Big(\sqrt{(u_i/\Delta u - q_{u, p})^2 + (v_i/\Delta u - q_{v, p})^2}\\, w_i - \bar{w}_k, \Delta u\Big)\, .
	\end{split}
\end{equation}
$(q_{u, p}, q_{v, p})$ represents the nearest grid points, and we use adaptive quadrature to calculate
\begin{equation}
	\begin{split}
		[GC]\Big(\sqrt{u_{\rm pix}^2 + v_{\rm pix}^2}, w, \Delta u\Big) =
		\frac{2\pi}{\Delta u ^2}\int_{0}^{\alpha/2} g(r)\\
		\times{\rm e}^{-2\pi iw(\sqrt{ 1 - r^2/\Delta u^2} - 1)}
		J_0\left(2\pi r \sqrt{u_{\rm pix}^2 + v_{\rm pix}^2}\right) r{\rm d}r\, ,
	\end{split}
	\label{eq:analytic_convolution_hankel}
\end{equation}
where $g(r)$ is the radial anti-aliasing filter in the image domain (i.e. the Fourier transform of the Kaiser-Bessel function), $\Delta u$ is the resolution of the Fourier grid as determined by the zero padded field of view, and $(u_{\rm pix},v_{\rm pix})$ are the pixel coordinates on the Fourier grid.

For each stack \mbox{$\bm{y}_k \in \mathbb{C}^{M_k}$} we have the measurement equation $\bm{y}_k = \bm{\mathsf{\Phi}}_k\bm{x}$. It is clear that each stack has an independent measurement equation. However, the full measurement operator is related to the stacks in the adjoint operators such that
\begin{equation}
	\bm{x}_{\rm dirty} =  \bm{{\rm AllSumAll}}_k\left(\bm{\mathsf{\Phi}}_k^\dagger \bm{y}_k\right) = \bm{\mathsf{\Phi}}^\dagger \bm{y}\, .
\end{equation}
We use an MPI all-sum-all to generate the same dirty map on each node. The full operator MPI $\bm{\mathsf{\Phi}}$ is normalized using the power method.
For further details see \citet{pra19b}.

\section{Bottleneck of the distributed stacking method}
\label{sec:bottleneck}
To minimize the time taken to perform kernel calculation and increase accuracy of the non-coplanar correction, the visibilities need to be sorted into $w$-stacks using a cluster algorithm. We do this by using the $k$-means clustering algorithm after using complex conjugation to reflect the visibilities to have positive $w$ \citep{pra19d}. Because the $w$-stacks are clustered to minimize error, the memory and computational load of each $\bm{\mathsf{[GC]}}_k$ has previously been ignored when assigning one stack $k$ per compute node. When the majority of visibilities lie in only a few stacks, the total available memory and resources for construction and application of $\bm{\mathsf{[GC]}}_k$ is bottlenecked. This is especially the case when there is one $\bm{\mathsf{[GC]}}_k$ per MPI node. This problem is emphasized for extremely wide-fields of view and large values of $w$, where the $w$-projection kernel size scales as $\frac{2w}{\Delta u}$, with $\frac{1}{\Delta u}\propto \text{field of view}$, and for large numbers of visibilities. Hence, these factors have a large impact on the required computational resources in kernel construction and application, as we demonstrate in Section \ref{sec:impl}.

In the next section we describe an algorithm that solves this bottleneck. We split the operator $\bm{\mathsf{[GC]}}_k$ into smaller operators $\bm{\mathsf{[GC]}}_{jk}$ that can be spread across multiple nodes $j$ for $w$-stacks indexed by $k$. We remove the requirement that image domain correction and Fourier domain correction are applied on the same node. We restrict the index $j$ for nodes that apply Fourier domain correction and index $k$ for nodes that apply image domain correction. This allows even distribution of the memory load, kernel construction, and application of the operator $\bm{\mathsf{[GC]}}$ to ensure scalability as demonstrated in Section \ref{sec:impl}.

\section{All-to-all distributed measurement operator}
\label{sec:mo_dist_grid}
In this section we introduce a new MPI distribution strategy for the application of a wide-field measurement operator. This process allows the FFTs of the $w$-stacks to be evenly distributed across all nodes while allowing the sparse matrix operations to be distributed evenly across all nodes. Communicating only the grid points that are needed for degridding minimizes communication in an intermediate all-to-all operation.

\subsection{Distributing measurements for computational load}
First the $k$-means algorithm is used to sort the visibilities into $w$-stacks $\bm{y}_k$. The visibilities of each stack $\bm{y}_k$ are distributed across MPI nodes $\bm{y}_{jk}$, where $1 \leq j \leq n_{\rm d}$, to evenly distribute the computation of $\bm{\mathsf{[GC]}}$. The computational load of an individual visibility $\bm{{y_\mathsf{k}}}_i$ is determined by the support size 
\begin{equation}
	{\rm support}(w_i - \bar{w}_k, \Delta u) = \max \{ J_{\rm min}, 2 (w_i - \bar{w}_k)/\Delta u\}\, ,
\end{equation}
where $J_{\rm min}$ is the 1d support size of the anti-aliasing kernel \citep{pra19b}. It is then straightforward to determine the total computational load of $\bm{\mathsf{[GC]}}$ and then distribute it evenly across nodes $j$. This is done by calculating the average computational load across all nodes from $j = 1$ to $j = n_{\rm d}$ in order, filling each node $j$ with visibilities until it reaches the average computational load. 

In practice, it is difficult to fill each node with the exact average computational load, because each visibility has its own integral (indivisible) computational load. This can be accommodated by allowing the last node to overfill slightly and keeping the rest of the nodes under the average load. Testing has shown that the overfill amount on the last node is insignificant.

\subsection{All-to-all distribution of Fourier grid subsections}
\begin{figure*}
\center
\includegraphics[width=12cm]{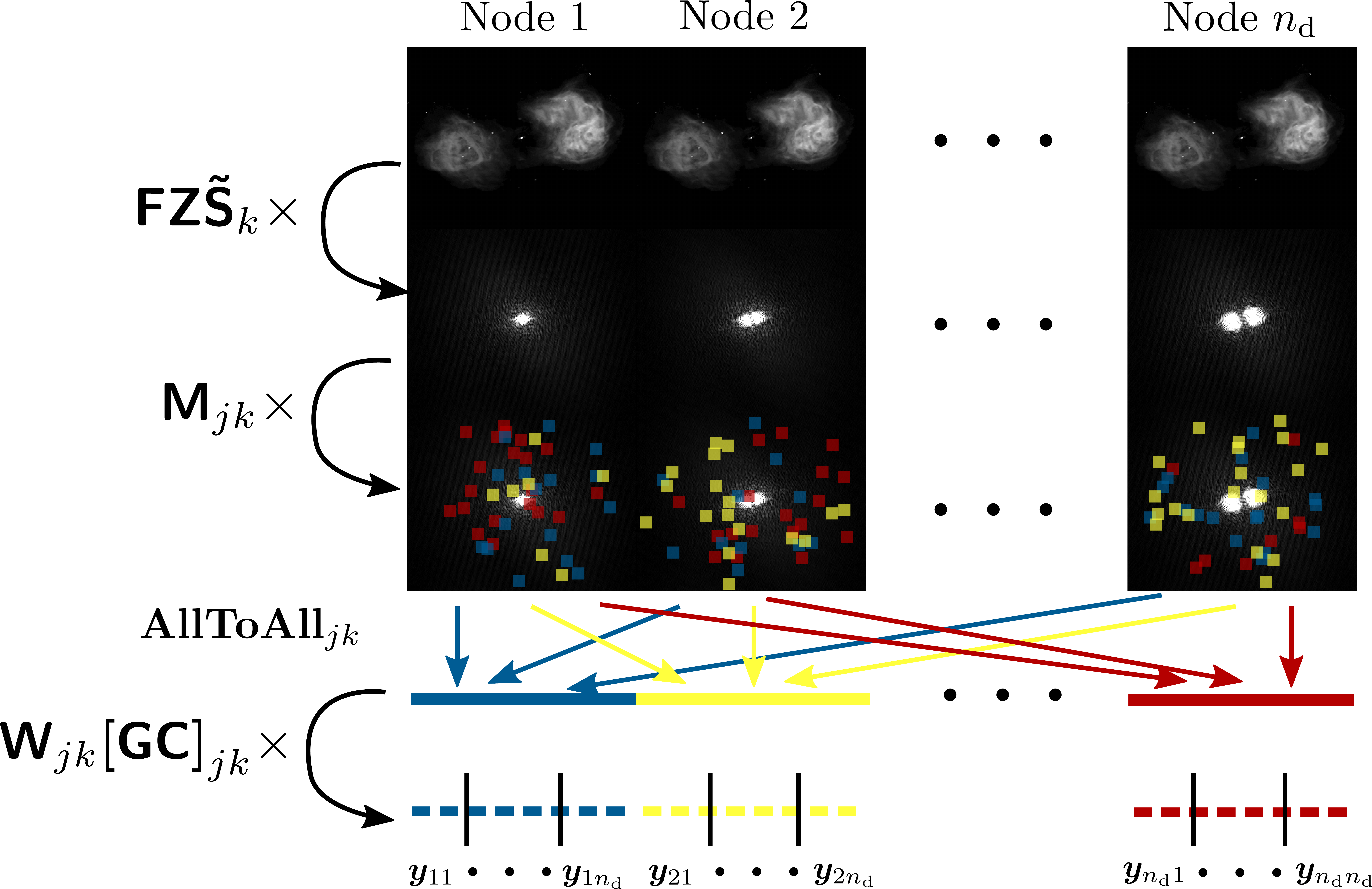}
\caption{Each node starts with a copy of $\bm{x}$. The linear operation $\bm{\mathsf{\tilde{S}}}_k$ applies the gridding correction and multiplication of the chirp on node $k$. Each node performs zero padding and an FFT with the operation $\bm{\mathsf{FZ}}$. The operation $\bm{\mathsf{M}}_{jk}$ selects sections of the FFT grid on node $k$ that are required on node $j$ for degridding (this is determined by the columns of $\bm{\mathsf{[GC]}}_{jk}$). The colored squares show regions of the grid that are to be sent to each node, with each color corresponding to a value of $j$. The sections of the FFT grid are distributed through a distributed MPI all-to-all communication step. This is followed by the application of $\bm{\mathsf{[GC]}}_{jk}$ for the $k^{\rm th}$ $w$-stack on node $j$, to interpolate the visibilities $\bm{y}_{jk}$ off of the grid, with the $w$-projection kernel performing the correction for the offset $w - \bar{w}_k$. The adjoint process corresponds to performing each step in reverse, followed by an all-sum-all operation over the $w$-stacks.}
\label{fig:algo_diagram}
\end{figure*}

With the computational load of $\bm{\mathsf{[GC]}}$ distributed across the nodes, the measurement equation needs to map sections of the grid that need to be sent to each node $j$ from each stack $k$ to minimize communication. Without loss of generality, we let $1 \leq k, j \leq n_{\rm d}$. The MPI measurement equation reads
\begin{equation}
	\bm{y}_{jk} = \bm{\mathsf{W}}_{jk}\bm{\mathsf{[GC]}}_{jk}
	\bm{{\rm AllToAll}}_{jk}\left(
	\bm{\mathsf{M}}_{jk}\bm{\mathsf{F}}\bm{\mathsf{Z}}\bm{\mathsf{\tilde{S}}}_k\bm{x}
	\right)\,,
\end{equation}
where the chirp multiplication and FFT are applied on node $k$ (assuming one $\bm{\mathsf{\tilde{S}_k}}$ per node for simplicity), the operator $\bm{\mathsf{M}}_{jk} \in \mathbb{R}^{K_{jk} \times K}$ selects only the grid sections (of size $K_j$) of the FFT grid (of size $K$) of stack $k$ that are needed for degridding on node $j$, which are then sent to node $j$ with the MPI all-to-all operation. This is followed by degridding to the visibilities on node $j$ that belong to stack $k$ using $\bm{\mathsf{[GC]}}_{jk} \in \mathbb{C}^{M_{jk}\times K_{jk}}$. In practice, $\bm{\mathsf{[GC]}}_{jk}$ are combined into one sparse matrix on each node that has $\sum_k M_{jk}$ rows and $\sum_j K_{jk}$ columns.
This entire process is visualized in Figure \ref{fig:algo_diagram}.

The application of the adjoint operator reads
\begin{equation}
\begin{split}
	\bm{x}_{\rm dirty} = \bm{{\rm AllSumAll}}_k\Big(
	\bm{\mathsf{\tilde{S}}}_k^\dagger\bm{\mathsf{Z}}^\dagger\bm{\mathsf{F}}^\dagger\times\quad\quad\quad\quad\quad\quad\quad\quad\\
	\sum_{j = 1}^{n_{\rm d}}\left [\bm{\mathsf{M}}_{jk}^\dagger
	\bm{{\rm AllToAll}}_{kj}\left(
	\bm{\mathsf{[GC]}}_{jk}^\dagger\bm{\mathsf{W}}_{jk}^\dagger \bm{y}_{jk}
	\right)\right]
	\Big)\,,
\end{split}
\end{equation}
where node $j$ grids visibilities from stack $k$, these grid sections are sent from node $j$ to stack $k$ through an all-to-all operation. The grid sections from each node $j$ are added to the full FFT grid of each stack $k$. An inverse FFT is applied followed by cropping of the image. Multiplication of the conjugate chirp is applied on each stack $k$ followed by an all-sum-all of the images to produce the same dirty map on each MPI node.

Extensive unit testing has shown that the distributed computation is equivalent to the non distributed computation and the standard $w$-stacking $w$-projection algorithm. It is worth noting that when $n_{\rm d} \times K > 2^{32} - 1$, 64 bit integer types are specifically needed for indexing across $n_{\rm d} \times K$ FFT pixels without overflow.

\section{Implementation and Application}
\label{sec:impl}
In this section we demonstrate the effectiveness of evenly distributing the computational load using the algorithm presented in Section \ref{sec:mo_dist_grid}. This algorithm has been implemented in the interferometric imaging software package PURIFY using C++ and MPI.  PURIFY is powered by distributed convex optimisation algorithms implemented in the software package SOPT\footnote{\url{https://github.com/astro-informatics/sopt}}.

To demonstrate the effectiveness of the algorithm, we simulate reconstruction of a 25 by 25 deg field of view, using a Gaussian variable sampling density in $uvw$ following \citet{LP18}. $u$ and $v$ are represented in radians, with a standard deviation of $\pi/3$. $w$ is represented in wavelengths, with a standard deviation of $200$ wavelengths, but was constrained to values between $\pm 600$ wavelengths. An 1024 by 1024 pixel image of M31 is considered, where the pixel size is 90 by 90 arcseconds. We add Gaussian noise to the measurements, so that the visibilities have an input signal to noise ratio of 30 decibels \cite{LP18}. We then apply the alternating direction method of multipliers (ADMM) algorithm as performed in \citet{LP18,pra19b,pra19c}. We used a minimal gridding kernel support size of $J_{\rm min} = 4$ for the Kaiser-Bessel kernel.

First we use conjugate symmetry to reflect the visibilities to have $w \geq 0$. Then we use the $k$-means clustering algorithm to assign each visibility to a $w$-stack indexed by $k$ and to calculate each $\bar{w}_k$. Then we iterate through the visibilities to assign the computational load across the nodes, following Section \ref{sec:mo_dist_grid}. The visibilities and $w$-stack indexes are redistributed using an all-to-all operation. Then the $w$-projection kernels shown in Equation \ref{eq:analytic_convolution_hankel} are constructed using adaptive quadrature to an accuracy of $10^{-6}$ in absolute and relative error, which has shown to be accurate to $1\%$ in the image domain \citep{pra19b}. This corrects each visibility for the $w$ offset determined by $\bar{w}_k$ and the $w$-stack index $k$.

We perform reconstruction using 2 billion visibilities with 50 nodes of the Grace supercomputing cluster at University College London (UCL). Each node has two 8 core Intel Xeon E5-2630v3 processors and 64 Gigabytes of RAM\footnote{More details can be found at \url{https://wiki.rc.ucl.ac.uk/wiki/RC_Systems\#Grace_technical_specs}}. Note that this is exactly the same configuration used in the recent work of \citealt{pra19d}, where an MWA Fornax A observation was reconstructed using 126 million visibilities.

The memory used to store $\bm{\mathsf{[GC]}}$ is distributed across 50 compute nodes. The memory needed to store $\bm{\mathsf{[GC]}}$ was approximately 21 Gigabytes on each node (3 Tb across all nodes). However, for efficient layout for memory access $\bm{\mathsf{[GC]}}^\dagger$ was also stored, requiring an additional 3 Tb across all nodes. The 2 billion visibilities amounts to 32 Gigabytes spread evenly across the nodes. To store the weights and $uvw$-coordinates during construction of $\bm{\mathsf{[GC]}}$ requires 64 Gigabytes of memory spread evenly over the cluster.

Sorting and distributing the visibilities took approximately 2 minutes. Kernel construction took 1 hour and 5 minutes. Application of the combined gridding and degridding operation took approximately 25 seconds. The ADMM algorithm converged in approximately 20 minutes with 9 iterations. The signal to noise ratio of the reconstruction was calculated as in \citet{LP18} to be 31.49 decibels.

Applying the standard distribution method of the $w$-stacking $w$-projection hybrid algorithm was not possible for the scenario considered due to memory requirements, where each $\bm{\mathsf{[GC]}}_k$ requires approximately 1 to 50 Gigabytes of memory. Additionally, even if there was enough memory on each node, run time would increase greatly due to lack of CPU cores on the heavily loaded nodes acting as a bottleneck. It is clear that the distribution method presented in this work circumvents this bottleneck in resources and enables accurate interferometric image reconstruction over extremely wide-fields of view for very large data sets.

\section*{Acknowledgments}
This work was supported by the Engineering and Physical Sciences Research Council (EPSRC) through grant EP/M011089/1.

\bibliographystyle{mymnras_eprint}
\bibliography{refs}



\label{lastpage}
\end{document}